\documentclass[useAMS,usegraphicx,usenatbib]{mn2e}

\usepackage{ulem}
\usepackage{color}
\usepackage{graphics,graphicx}
\usepackage{times} 
\usepackage{amssymb}
\usepackage{amsmath}
\usepackage{lscape}
\usepackage{url}
\usepackage{multirow}
\usepackage{ctable}
\usepackage{threeparttable}
\newif\ifAMStwofonts
\AMStwofontstrue
\title{Comparing Spectral Models for ULXs with NGC\thinspace 4517 ULX1}

\author[D.\,J. Walton et al.]
{\parbox{7.in}{D.\,J. Walton$^{1}$ \thanks{E-mail: dwalton@ast.cam.ac.uk},
J.\,C. Gladstone$^{2}$, 
T.\,P. Roberts$^{3}$, 
A.\,C. Fabian$^{1}$,
M.\,D. Caballero-Garcia$^{1}$ \\
C. Done$^{3}$ \& 
M.\,J. Middleton$^{3}$ \\ 
\footnotesize
$^1$ \it{Institute of Astronomy, Cambridge University, Madingley Road, Cambridge, CB3 0HA} \\ 
$^2$ \it{Department of Physics, University of Alberta, Edmonton, Alberta, T6G 2C7, Canada} \\
$^3$ \it{Department of Physics, University of Durham, South Road, Durham, DH1 3LE}}}
\date{}

\def\ein{{\it Einstein~\/}}
\def\xmm{{\it XMM-Newton~\/}}

\def\suzaku{{\it Suzaku}}

\def\swift{{\it Swift}}

\def\nustar{{\it NuSTAR}}

\def\dss{{\it DSS~\/}}
\def\epic{{\it EPIC~\/}}
\def\epicpn{{\it EPIC}{\rm-pn}}
\def\epicmos1{{\it EPIC}{\rm-MOS1~\/}}
\def\epicmos2{{\it EPIC}{\rm-MOS2 ~\/}}
\def\epicmos{{\it EPIC}{\rm-MOS}}

\def\ned{the NASA Extragalactic Database}

\def\ks{\hbox{$\rm\thinspace ks$}}

\def\deg{$^{\circ}$}

\def\mpc{\hbox{$\rm\thinspace Mpc$}}

\def\as{\hbox{$\rm\thinspace arcsec$}}

\def\H0{{\rm km~s$^{-1}$~Mpc$^{-1}$}}

\def\kev{\hbox{$\rm\thinspace keV$}}

\def\ergsec{{\rm ~erg~s$^{-1}$}}

\def\atpcm{{atom~cm$^{-2}$}}

\def\ergps{\hbox{erg~s$^{-1}$}}
\def\ergcmps{\hbox{\rm erg~cm~s$^{-1}$}}

\def\msun{\hbox{$\rm\thinspace M_{\odot}$}}

\def\chisq{{$\chi^{2}$}}
\def\rchi{{$\chi^{2}_{\nu}$}}

\def\pl{\rm{\small POWERLAW}}

\def\phabs{\rm{\small PHABS~\/}}
\def\zphabs{\rm{\small ZPHABS~\/}}
\def\diskbb{\rm{\small DISKBB~\/}}
\def\dkbbfth{\rm{\small DKBBFTH}}

\def\refhiden{\rm{\small REFHIDEN}}

\def\reflionx{\rm{\small REFLIONX}}

\def\kdblur{\rm{\small KDBLUR}}
\def\kdblurtwo{\rm{\small KDBLUR2}}

\def\pexrav{\rm{\small PEXRAV}}

\def\xspecv{\hbox{\small XSPEC}\thinspace v12.5.1n\thinspace}

\def\xmmselect{\hbox{\rm{\small XMMSELECT}}}
\def\ftool{\hbox{\rm{\small FTOOL}}}

\def\addspec{\hbox{\rm{\small ADDSPEC}}}

\def\grppha{\hbox{\rm{\small GRPPHA~\/}}}

\def\sas{\hbox{\rm{\small SAS~\/}}}
\def\epchain{\hbox{\rm{\small EPCHAIN~\/}}}
\def\emchain{\hbox{\rm{\small EMCHAIN~\/}}}

\def\rmfgen{\hbox{\rm{\small RMFGEN}}}
\def\arfgen{\hbox{\rm{\small ARFGEN}}}
\def\epiclccorr{\hbox{\rm{\small EPICLCCORR}}}

\def\grid25{\hbox{\rm{\small GRID25}}}

\def\ka{K~$\alpha$}

\def\etal{et al.~\/}
\def\eg{{\it e.g.~\/}}
\def\etc{{\it etc.}}
\def\ie{{\it i.e.~\/}}

\def\la{\mathrel{\hbox{\rlap{\hbox{\lower4pt\hbox{$\sim$}}}{\raise2pt\hbox{$<$}}}}}
\def\ga{\mathrel{\hbox{\rlap{\hbox{\lower4pt\hbox{$\sim$}}}{\raise2pt\hbox{$>$}}}}}

\def\d25{D$_{25}$}
\def\nh{{$N_{\rm H}$}}

\def\.25{0.25 keV\thinspace}
\def\lx{$L_{\rm X}$}

\def\mbh{\rm $M_{\rm BH}$}

\def\rg{$R_{\rm G}$}
\def\rin{$R_{\rm in}$}
\def\rout{$R_{\rm out}$}

\def\ion{$\xi$}

\def\mdot{$\dot{M}$}
\def\lx{$L_{\rm X}$}

\def\ulxone{\rm{NGC\thinspace4517 ULX1}}
\def\ngc4517{\rm{NGC\thinspace4517}}

\begin{document}
\pagerange{\pageref{firstpage}--\pageref{lastpage}}
\pubyear{2009}
\maketitle
\label{firstpage}

\begin{abstract}
We present the previously unanalysed high quality \xmm spectrum of an ultraluminous
X-ray source candidate in \ngc4517. As with other high quality ULX spectra, a
downturn in the spectrum is observed at $\sim$6\,\kev. Both of the recent disc
reflection and Comptonisation interpretations of this feature are applied, in order
to present a direct comparison, and are found to provide statistically equivalent
representations of the current data. We find that the reflection model requires the accretion
disc to have a highly super-solar iron abundance, while the Comptonisation model
requires low temperature Comptonising electrons, and for the corona to be optically
thick. These physical requirements are discussed in detail, and physically motivated scenarios are
highlighted in which each model can be considered a viable explanation for the observed
emission. By extending our consideration of these two interpretations to high energies,
we demonstrate that observations of ULXs at energies $\gtrsim10$\kev\ should be extremely
useful when attempting to distinguish between them. With current instrumentation, it is
only viable to perform these observations for M~82 X-1, but future
high angular resolution hard X-ray imaging spectrometers, such as the Hard X-ray Imaging
System due to fly on \textit{Astro-H}, should go a long way to resolving this issue.
\end{abstract}

\begin{keywords}
X-rays: binaries -- black hole physics
\end{keywords}

\section{Introduction}

Ultraluminous X-ray Sources (ULXs), discovered with the \ein observatory in
the 1980s (\citealt{Fabbiano89}), are extra-nuclear point sources observed
to be more luminous in X-rays than the Eddington luminosity for a stellar
mass ($\sim$10\msun) black hole, \ie $L_{X}>10^{39}$\ergsec, a combination
which has led to extended debate over the nature of these sources. The
extremity of the observed luminosities implies that these sources are most
likely to be accreting black holes of some kind.

There are currently three main theories to explain the apparent X-ray
luminosities of ULXs. The first is that these sources are systems similar in
nature to the less luminous X-ray binaries (XRBs) frequently observed, but
in which the mass of the black hole is greater than the stellar-mass black
holes that XRBs are expected to harbour. However, they must also be less
massive than the supermassive black holes associated with active galactic
nuclei (\mbh$ >10^{5-6}$\msun; SMBH), as dynamical friction dictates that
such objects would sink to the centres of their respective galaxies well
within a Hubble Time (\citealt{MilCol04}). In this case, ULXs are interpreted
as intermediate mass black holes with $10^{2}\msun \lesssim$ \mbh\ $\lesssim
10^{6}\msun$ (IMBH; \citealt{Colbert99}). The second is that ULXs are actually
the same systems as regular XRBs, but observed in a different accretion state
in which the black hole is able to radiate at super-Eddington rates (for a
recent review of the standard accretion states observed in XRBs see
\citealt{Remillard06}), and a number of methods have been proposed by which
this might be possible (see \eg \citealt{Poutanen07}, \citealt{Finke07}).

The third again involves stellar-mass XRBs, but in this case the observed
emission is not isotropic (\citealt{King01}). ULXs may be such sources with
which we have a favourable orientation, and as such artificially high
luminosities are calculated for these sources based on the assumption of
isotropic emission. Violation of the Eddington limit may be avoided by
assuming a suitably high level of anisotropy. Such anisotropy could be due
to geometrically thick accretion discs subtending a larger solid angle,
funnelling the X-rays produced in the inner regions. Such `slim' discs may
also allow a compact object to radiate at super-Eddington rates
(\citealt{Abram80}), so super-Eddington and anisotropic emission may be
intrinsically linked. Alternatively anisotropic emission could manifest
itself in the form of pencil-beam relativistic jets similar to those seen
in other accreting sources (\citealt{Reynolds97}). However, a number of
ULXs have been observed to be embedded within (roughly) spherically symmetric
photoionised emission nebulae inconsistent with narrow beaming (\eg
\citealt{Pakull03}, \citealt{Kaaret04}, \citealt{Berghea10}). This suggests
highly anisotropic emission by itself seems unlikely to be able to explain
ULXs as a class, although it cannot be ruled out in individual sources. For
a more detailed review on the possible nature of ULXs see \cite{Roberts07}.

Observationally it has been difficult to distinguish between the IMBH and
stellar mass interpretations. ULXs are extragalactic sources hence
detailed information on optical counterparts is still relatively rare,
although this is being addressed, and orbital periods/parameters of the
assumed binary system are even rarer still; attempts at dynamical
measurements of the black hole mass are only now becoming possible for a
small number of sources with the use of world leading observational
facilities. Current examples of these attempts are limited to periodicity
claims in M~82 X-1 (\citealt{Kaaret07}), NGC~1313 X-2 (\citealt{Liu09}) and
NGC~5408 X-1 (\citealt{Strohmayer09}), although in the latter case
\cite{Foster10} suggest the periodicity might be superorbital, related to
jet precession. Much recent work has been focused on successfully modelling
the X-ray spectral and timing properties of ULXs, as both the inner
temperature of the accretion disc and characteristic timescales are mass
dependent, proportional to $M_{\rm BH}^{-1/4}$ and $M_{\rm BH}^{-1}$
respectively (\citealt{Makishima00}, \citealt{Vaug03}). Early results
pointed towards an IMBH interpretation with modelled disc temperatures
being cooler and characteristic timescales being longer than XRBs (see \eg
\citealt{MilFab04, Strohmayer07}). However, a systematic study of the
highest quality ULX X-ray spectra by \cite{Stobbart06} showed the majority
displayed a break/turnover at high energies ($\gtrsim3$\kev). Such breaks
are not commonly seen in the spectra of XRBs.

\begin{figure}
\begin{center}
\rotatebox{0}{
{\includegraphics[width=235pt]{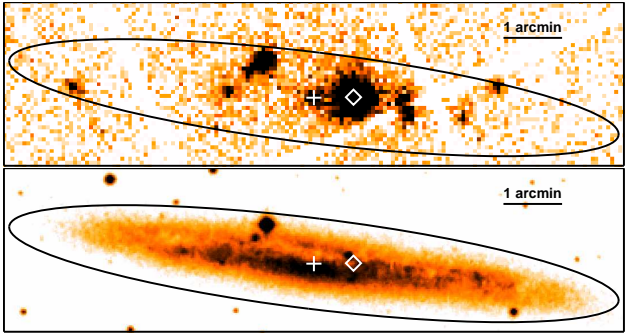}}
}
\end{center}
\caption{\xmm X-ray (top) and \dss optical (bottom) images of NGC 4517. The
elliptical region shows the RC3 D25 isophote, and the cross and diamond
indicate the positions of the nucleus and \ulxone\ respectively as the
position error circles for the X-ray detections are too small to show.}
\label{fig_pos}
\end{figure}

\cite{Gladstone09} argue that this turnover could be due to Comptonisation
from an optically thick corona, which would shroud the inner regions of
the accretion disc and artificially lower the inner temperatures obtained
from simple spectral modelling. Under this assumption, the `true' inner
disc temperatures recovered are akin to those observed in XRBs. Most ULXs
would then represent a new, high Eddington fraction accretion state for
these stellar-mass black holes. An alternative explanation is proposed by
\cite{Caball10} who note that in many cases the turnover occurs between
$\sim$5-7\kev. They demonstrate it could be due to a combination of the iron
\ka\ emission line and absorption edge (6.4 and 7.1 \kev\ respectively) in a
relativistically blurred reflection spectrum from the inner accretion disc
of a spinning black hole. Here `reflection' refers to the backscattering
and fluorescence of X-rays (\citealt{George91}). In this case the excess
soft emission often assumed to come directly from the accretion disc is
also due to blurred reflection, so the disc is not actually observed, hence
the inner temperature cannot be measured and the mass must be estimated by
some other method. This explanation for the soft emission is conceptually
similar to that proposed by \cite{Goncalves06}. Other explanations have also
been proposed, often based on emission from the aforementioned slim discs in
which advection of radiation from the inner disc reduces the observed
luminosity from this region (\citealt{Abram88}), although these have not been
as successful in reproducing the observed spectra (see \eg attempts in
\citealt{Gladstone09}).

It is important to determine the origins of this curvature, as it is one of the
properties of ULXs that distinguishes them from their less luminous XRB cousins.
Here we present an \xmm observation of a new ULX candidate located in NGC 4517
(hereafter \ulxone), which was brought to light during work on a new catalogue
of ULX candidates in nearby galaxies (Walton \etal 2011, submitted). Both
reflection and Comptonisation interpretations are applied during the spectral
analysis and the models directly compared. The paper is structured as follows:
section \ref{sec_red} details the data reduction, section \ref{sec_spec}
describes the spectral analysis, section \ref{sec_dis} discusses and compares
the two models, focusing on how they may be distinguished, and finally section
\ref{sec_conc} presents our conclusions.

\section{Observations and Data Reduction}
\label{sec_red}

\subsection{Host Galaxy}
\label{subsec_host}

NGC~4517 (also known as NGC~4437) is an Sc type spiral galaxy located at RA =
$12^{h}32^{m}45.6^{s}$, Dec = $+00$\deg$06'54.0''$ with a distance of $\sim$18
\mpc\ (assuming $H_0 = 73$ \H0, $\Omega_{\rm matter} = 0.27$, $\Omega_{\rm vacuum}
= 0.73$; the redshift of NGC~4517 is $z=0.003764$)\footnote{The basic
information for NGC 4517 presented here has been obtained from \ned\ (NED);
http://nedwww.ipac.caltech.edu/}, and has an apparent edge-on orientation.
Although it has received relatively little observational attention, its
orientation has lead to NGC 4517 being a popular galaxy to include in studies
of Globular Clusters (\eg \citealt{Goudfrooij03} and \citealt{Chandar04}),
potentially important environments for the formation of IMBHs (see
\citealt{Noyola08}, and references therein) and hence possibly ULXs. Based on
the diagnostic system of \cite{Kewley06}, \cite{Seth08} find nuclear emission line
ratios consistent with star formation being a significant source of energy within
NGC\,4517. Star forming regions are also important environments for ULX formation
(\citealt{Swartz09}). However, they also find evidence of AGN activity, and do not
go as far as to provide a quantitative estimate for the star formation rate.

The only targeted X-ray observation of NGC\,4517 was taken in December 2004 when
it was observed with \xmm (\citealt{XMM}) for $\sim$114 \ks, during which a number
of X-ray point sources were detected within its $D_{25}$ isophote, as defined in
the RC3 galaxy catalogue (\citealt{RC3}), including a point source at RA =
$12^{h}32^{m}45.0^{s}$, Dec = $+00$\deg$06'55.0''$, co-incident with the expected
location of the galaxy nucleus. However, the brightest of these sources, with an
observed (absorbed) X-ray luminosity of $L_{0.2-12.0} \sim 2 \times 10^{40}$\ergsec\
(assuming isotropic emission and that the association with NGC\,4517 is correct),
is \ulxone\ (IAU ID: 2XMM J123242.7+000654), offset from the nucleus by 43\as\ (see
Fig. \ref{fig_pos}), with position RA = $12^{h}32^{m}43.0^{s}$, Dec =
$+00$\deg$06'55.0''$. It is this source that is the subject of this work. In contrast,
the X-ray detection coincident with the galactic centre is observed to have a
luminosity of only $L_{0.2-12.0}\sim 10^{39}$ \ergps. The X-ray detections of the ULX
candidate and the nuclear source have position errors of 1.1'' and 1.8'' respectively
(99 per cent confidence level), as quoted in the 2XMM Serendipitous Survey
(\citealt{2XMM}), while the NED position for the centre of NGC~4517 should be
accurate to within 3.8'', so we may be confident \ulxone\ is not associated with the
nucleus of its host galaxy.

\subsection{Data Reduction}

\begin{figure}
\begin{center}
\rotatebox{270}{
{\includegraphics[width=155pt]{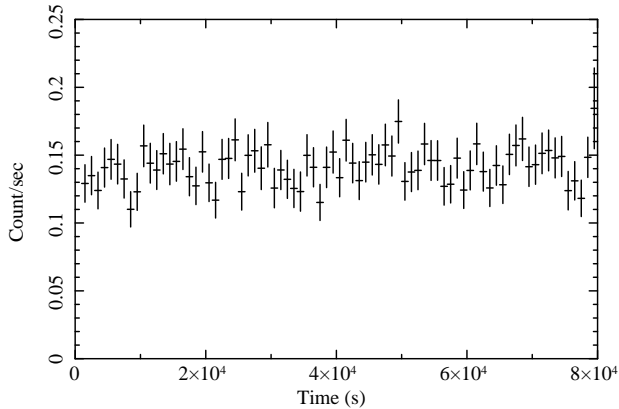}}
}
\end{center}
\caption{The \epicpn\ lightcurve of \ulxone\ with 1000\,s resolution, obtained during
the \xmm observation of NGC 4517 taken in December 2004.}
\label{fig_lc}
\end{figure}

The data reduction was carried out with the Science Analysis System (\sas v8.0.1)
largely according to the standard prescription provided in the online
guide\footnote{http://xmm.esac.esa.int/}. The observation data files were processed
using \epchain and \emchain to produce calibrated event lists for the \epicpn\
(\citealt{XMM_PN}) and \epicmos\ (\citealt{XMM_MOS}) CCDs respectively. Lightcurves
and spectra were produced for the 0.3--10.0\,\kev\ energy range selecting only
single and double events (single to quadruple events) for \epicpn\ (\epicmos) using
\xmmselect, and periods of high background were treated according to the method
outlined by \cite{Picon04} with the signal-to-noise ratio maximised for the full
considered energy band. A circular region of radius 29''/6.7 pix. centred on \ulxone\
was defined in order to include as many source counts as possible without
contamination from the nearby nuclear source or including CCD chip edges. The
observation was targeted at the galaxy position, so \ulxone\ was offset from the
nominal pointing by $\sim$45'' for all detectors. A second, larger circular region
of radius 99''/22.7 pix. was chosen in an area of the same CCD free of other sources
to sample the background. We also extracted the spectrum of the source
coincident with the nucleus, using a circular region of radius 16''/3.7 pix. so as
not to overlap the region adopted for the ULX. The redistribution matrices and
auxiliary response files were generated with \rmfgen\ and \arfgen. Lightcurves were
corrected for the background count rate using \epiclccorr\footnote{Note that \epiclccorr\
only accounts for the instrumental good time interval (GTI) generated during the
production of the events file, it does not account for additional GTI files based
on \eg periods of high background, as used here.}. After performing the data reduction
separately for each of the MOS CCDs, the spectra were combined using the
\ftool\footnote{http://heasarc.nasa.gov/ftools/ftools\_menu.html} \addspec\
to improve the statistics at the highest and lowest energies. \addspec\ combines
spectra in a response weighted manner, to account for any differences there
may be in \eg the extraction regions \etc, and also automatically combines
the instrumental responses and background spectra associated with the source
spectra. Finally, spectra were re-binned using \grppha to have a minimum of 25
counts in each energy bin, so that the probability distribution of counts within
each bin can be considered Gaussian, and hence the use of the $\chi^2$
statistic is appropriate when performing spectral fits.

This procedure yielded good quality spectra for \ulxone, comparable with some of
the best ULX spectra currently obtained. In total, $\sim$15500 counts were
recorded with the \epic cameras, with $\sim$8500 of these recorded by \epicpn,
so the data meet the quality criterion adopted by both \cite{Gladstone09} and
\cite{Stobbart06}. The total good observation times contributing to the reduced
spectra are $\sim$70\,ks for \epicpn\ and $\sim$85\,ks for each \epicmos\ CCD. The
\epicpn\ lightcurve obtained for the first 80 ks of the observation (of which
$\sim$70\,ks has a low enough background count to be considered good time and
contribute to the spectrum) is shown in Fig. \ref{fig_lc}, during which the source
remains fairly constant. After this time, the observation is dominated by very
strong background flaring events. We attempt to calculate the fractional excess
variability (\citealt{Edelson02}), but find that the lightcurve is completely
dominated by statistical fluctuations and does not display any observable
intrinsic variability, similar to a number of other ULXs highlighted in
\cite{Heil09}.

\begin{table}
  \caption{A comparison of the single and broken powerlaw models, modified by
neutral absorption with column density $N_H \simeq 8\times10^{21}$ \atpcm, applied
to the 2-10\kev\ data for \ulxone\ (see text).}
\begin{center}
\begin{tabular}{c c c c c}
\hline
\hline
\\[-0.3cm]
Model & $\Gamma_{1}$ & $E_{\rm br}$ & $\Gamma_{2}$ & \rchi$(\mathrm{d.o.f})$ \\
& & (\kev) & & \\
\\[-0.3cm]
\hline
\\[-0.25cm]
\pl & $2.20 \pm 0.05$ & - & - & 1.4(268)\\
\\[-0.2cm]
\small{BROKEN} & \multirow{2}{*}{$1.97^{+0.07}_{-0.06}$} & \multirow{2}{*}{$5.5^{+0.7}_{-0.3}$} & \multirow{2}{*}{$3.9^{+1.3}_{-0.5}$} & \multirow{2}{*}{1.1(266)} \\
\\[-0.4cm]
\pl \\
\\[-0.3cm]
\hline
\hline
\end{tabular}
\label{tab_turn}
\end{center}
\end{table}

\section{Spectral Analysis}
\label{sec_spec}

Here we present our spectral analysis of \ulxone. Throughout this work, spectral
modelling is performed with \xspecv (\citealt{XSPEC}), and parameters are obtained
by modelling the \epicpn\ and combined \epicmos\ spectra simultaneously. All
parameters are tied between the two spectra with the exception of a constant
multiplicative component, required to be 1.0 for \epicpn\ but allowed to vary for
\epicmos\ to account for possible cross-calibration uncertainties. In all cases
presented in the following sections this parameter is within 5 per cent of unity.
All quoted uncertainties are the 90 per cent confidence limits for a single parameter
of interest, unless stated otherwise. The Galactic absorption column in the direction
of NGC~4517 is $\sim$$1.88 \times 10^{20}$ \atpcm\ (\citealt{NH}).

\subsection{High Energy Turnover}
\label{subsec_spec_turn}

\begin{figure*}
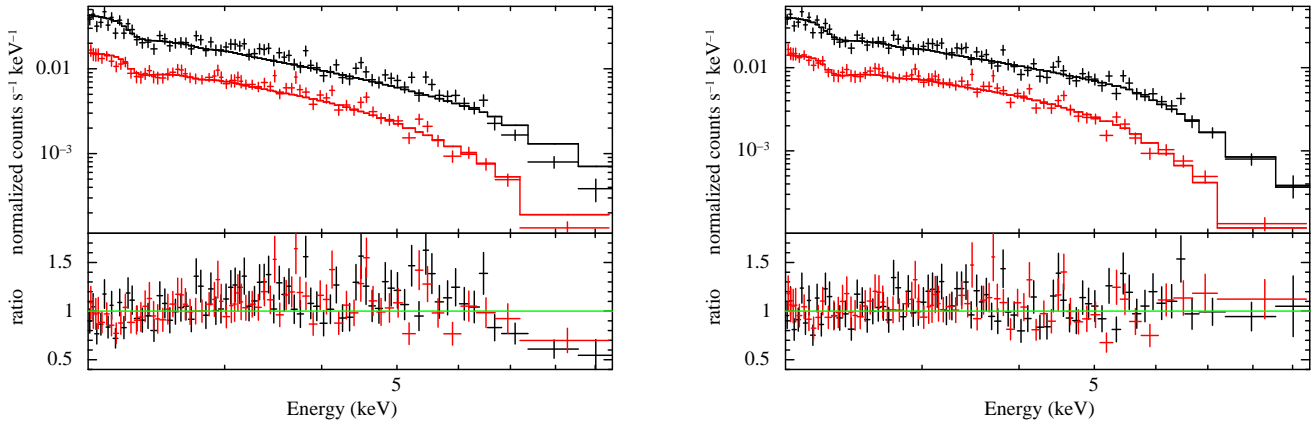

\begin{center}
\rotatebox{270}{
{\includegraphics[width=158pt]{./figs/NGC4517_ULX1_sp_turn_po.ps}}
}
\rotatebox{270}{
{\includegraphics[width=158pt]{./figs/NGC4517_ULX1_sp_turn_bknpo.ps}}
}
\end{center}
\caption{The \epicpn\ (black) and combined \epicmos\ (red) 2-10\kev\ spectra of \ulxone,
modelled with absorbed single and broken-powerlaw models (left and right panels
respectively, see text); data/model ratio plots are also shown for each. The single
powerlaw model clearly predicts an excess of counts over those observed at high energies
(above $\sim$6\,\kev). The data have been rebinned for display purposes only.}
\label{fig_turn}
\end{figure*}

As previously stated, high energy curvature is frequently seen in the best quality
X-ray spectra of ULXs (\citealt{Gladstone09}; \citealt{Stobbart06}). Given that the data
presented here is of similar quality, it is natural perhaps to expect that such a
turnover should also be observed in \ulxone. Accordingly, we investigated the high
energy data to determine whether a turnover was present by comparing single and
broken powerlaw models to the 2-10\kev\ spectra, following a similar prescription
to \cite{Stobbart06}. However, inspection of the full 0.3--10.0\,\kev\ data for \ulxone\
suggests there is a significant amount of neutral material obscuring the source,
with a column density of \nh\ $\simeq 8\times10^{21}$ \atpcm\ (far in excess of the
Galactic column). This is most likely due to the edge-on orientation of the host galaxy,
and unfortunately will add to the difficulty in reliably determining the presence and
origins of any excess emission at soft energies, and hence to the difficulty in
distinguishing between spectral models with the current data. Modelling the
(admittedly very poor) spectrum of the source associated with the nucleus with a
simple absorbed powerlaw also implies significant absorption, with a column density
of \nh\,$\simeq$\,$2\times10^{22}$\,\atpcm.

\begin{figure}
\begin{center}
\rotatebox{270}{
{\includegraphics[width=160pt]{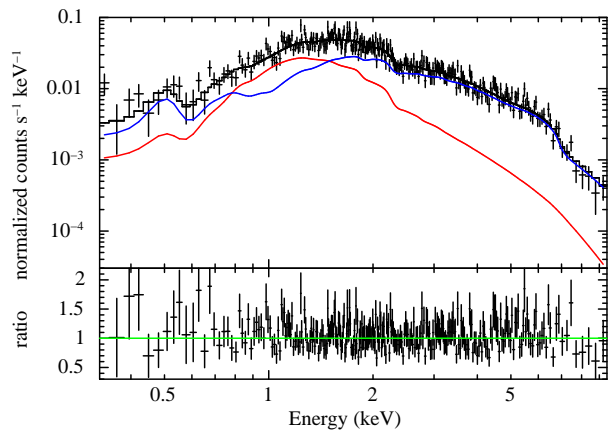}}
}
\end{center}
\caption{The disc reflection model applied to the spectrum of \ulxone\
and the relative contributions of the components included; the
solid black line is the total model, the red shows the powerlaw continuum and
the blue shows the blurred reflection component. Only the \epicpn\ data are shown for clarity.}
\label{fig_refl}
\end{figure}

While \cite{Stobbart06} compare unabsorbed models, the column required here will
have some effect on the spectrum above 2\,\kev, so we include absorption with a
column density \nh\ $\simeq 8\times10^{21}$ \atpcm\ in our comparison. We find that
the broken powerlaw model is significantly preferred, providing an improvement of
$\Delta$\chisq = 77 for 2 additional degrees of freedom (d.o.f.), the probability of
chance improvement given by an F-test is negligible, at $2\times10^{-14}$. The
comparison is shown in Table \ref{tab_turn} and Fig. \ref{fig_turn}, in which it
is clear there are fewer counts observed at high energies than predicted with
the single powerlaw model. This implies that the high energy spectrum of \ulxone\
does indeed display curvature, similar to that seen in other high quality ULX data.
In addition, the break energy of 5.5~\kev\ obtained is similar to that commonly
seen in the other high quality ULX spectra, where the break energy is often close
to $\sim$6~\kev.

Having demonstrated the presence of intrinsic high energy curvature, in the
following sections we apply the physically motivated disc reflection and
Comptonisation interpretations proposed to explain this feature. We do not
preclude that other interpretations for this source are possible, but we
limit ourselves to these two as a demonstrative process in order to
investigate how they may be distinguished in general.

\subsection{Disc Reflection Interpretation}
\label{subsec_spec_refl}

\begin{figure*}
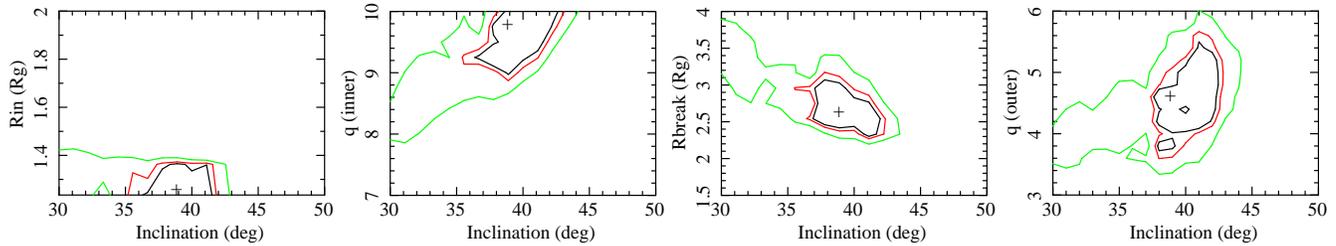

\begin{center}
\rotatebox{270}{
{\includegraphics[width=91pt]{./figs/contour_i_Rin.ps}}
}
\rotatebox{270}{
{\includegraphics[width=91pt]{./figs/contour_i_q1.ps}}
}
\rotatebox{270}{
{\includegraphics[width=91pt]{./figs/contour_i_Rbr.ps}}
}
\rotatebox{270}{
{\includegraphics[width=91pt]{./figs/contour_i_q2.ps}}
}
\end{center}
\caption{Confidence contours of the inclination paired with the other blurring
parameters: \rin,\,\ $q_{\rm inner}$,\,\ $R_{\rm br}$ and $q_{\rm outer}$ (left to right).
The contours shown are the 90, 95 and 99 per cent confidence levels for two
parameters of interest (black, red and green respectively). In each case the
best fit parameter values given in Table \ref{tab_refl} are marked with a +
symbol. There do not seem to be large parameter degeneracies present in the
reflection interpretation presented for \ulxone\ at even the 95 per cent level.}
\label{fig_cont}
\end{figure*}

We begin by considering the reflection model presented in \cite{Caball10}, in which
the observed turnover is due to the combined presence of a relativistically blurred iron
emission line and absorption edge. These features arise from X-ray reflection within the
inner regions of the accretion disc after irradiation of this material by the intrinsic
powerlaw continuum. This model makes use of the \reflionx\ grid (\citealt{reflion}), a
self consistent reflection model which intrinsically includes iron K-shell absorption
and emission. Its key parameters are the iron abundance of the reflecting medium (relative
to the solar iron abundance), $A_{\rm Fe}$, the photon index of the assumed ionising powerlaw
continuum, $\Gamma$, and the ionisation parameter of the surface of the reflecting medium,
$\xi=L/nR^{2}$, where $L$ is the incident luminosity able to ionise hydrogen, $n$ is the
number density of hydrogen and $R$ is the distance from the ionising source. \reflionx\
does not include the abundances of any elements other than iron as free parameters,
the model is instead calculated assuming their solar values as given in \cite{Morrison83}.
The relativistic blurring is applied using the convolution model \kdblur\ which accounts
for the extreme gravitational blurring expected in the innermost regions around a Kerr
black hole using calculations performed by \cite{kdblur}. Here the key parameters are the
inner and outer radii of the accretion disc, \rin\ and \rout\ respectively, its
inclination with respect to the observer, $i$, and its emissivity index, $q$,
which describes the assumed powerlaw emissivity profile of the disc,
of form $\epsilon(r)\propto r^{-q}$. Higher values of $q$ imply preferential
illumination of the inner regions of the disc, and hence a more compact, centrally
located corona (we caution the reader that although there are differences in
the Compton scattering medium between the reflection and Comptonisation
interpretations, in terms of the physical condition and possibly also the origin
of the matter, as will be discussed later, we refer to it as the corona in both cases).

Initially we applied the model following the same method as \cite{Caball10} including
both Galactic and intrinsic neutral absorption using \phabs and \zphabs respectively.
A reasonable fit with \rchi$(\mathrm{d.o.f})$ = 1.2(468) is obtained with similar
parameters to those presented in the initial paper, \ie significant blurring of the
reflected emission, with the accretion disc extending to the last stable orbit of a
maximally rotating black hole. The disc is required to have a highly super-solar iron
abundance, and the ionising continuum  is extremely steep in this case. However, the
turnover was still not adequately modelled. To resolve this, we make a minor
modification to the model (without changing the physical components included) and
replace \kdblur\ with \kdblurtwo, improving the fit to \rchi$(\mathrm{d.o.f})$ =
1.1(466) ($\Delta$\chisq of 40 for 2 extra d.o.f.). \kdblurtwo\ applies the
relativistic effects due to strong gravity in the same way as \kdblur, but allows the
emissivity profile of the disc $\epsilon(r)$ to have a broken powerlaw form, breaking
at $R_{\rm br}$. The parameters obtained with this modification are given in Table
\ref{tab_refl}, and the relative contributions of the components are shown in Fig.
\ref{fig_refl}. We still find the emissivity profile to be strongly centrally peaked.
During the application of the reflection model, the outer radius of the disc was set
to 400~\rg, the maximum allowed by the \kdblur\ and \kdblurtwo\ models, as it is very
poorly constrained due to the centrally peaked emissivity index. The only constraint
obtained if this parameter is allowed to vary is that the outer radius of the disc
must be greater than 10 \rg.

The disc reflection interpretation constructed is a complex, multi-parameter model.
We therefore investigated the confidence contours for various combinations of parameters
to search for parameter degeneracies, focussing on those of the \kdblurtwo\ component.
Degeneracies in the blurring parameters have been found to be a significant issue in a
number of cases during the application of reflection models to other sources (see \eg
\citealt{Nardini10}). Fig. \ref{fig_cont} shows the confidence contours of the disc
inclination paired with each of the remaining \kdblurtwo\ parameters (excluding \rout,
which was not free to vary). It is clear from these panels that strong parameter
degeneracies are not an issue for the reflection interpretation of this particular
source, even at the 95 per cent confidence level for two parameters of interest.

\begin{table}
  \caption{Parameters obtained with the reflection interpretation for the spectrum of NGC 4517
ULX1; parameters marked with * have not been allowed to vary.}
\begin{center}
\begin{tabular}{c p{0.2cm} c p{0.2cm} c}
\hline
\hline
\\[-0.3cm]
Component & & Parameter & & Value \\
\\[-0.3cm]
\hline
\hline
\\[-0.3cm]
\phabs & & \nh\tmark[a] & & $0.188$* \\
\\[-0.3cm]
\hline
\\[-0.3cm]
\zphabs & & \nh\tmark[a] & & $8.74^{+0.29}_{-0.65}$ \\
\\[-0.2cm]
& & $z$ & & $0.003764$* \\
\\[-0.3cm]
\hline
\\[-0.3cm]
\pl & & $\Gamma$ & & $>3.18$ \\
\\[-0.3cm]
\hline
\\[-0.3cm]
\kdblurtwo & & \rin\tmark[b] & & $1.242^{+0.065}_{-0.007}$ \\
\\[-0.2cm]
& & \rout\tmark[b] & & $400$* \\
\\[-0.2cm]
& & $i$ & & $38.6^{+3.2}_{-1.4}$ \\
\\[-0.2cm]
& & $q_{\rm inner}$ & & $>9.18$ \\
\\[-0.2cm]
& & $q_{\rm outer}$ & & $4.56^{+0.65}_{-0.59}$ \\
\\[-0.2cm]
& & $R_{br}$\tmark[b] & & $2.59^{+0.16}_{-0.35}$ \\
\\[-0.3cm]
\hline
\\[-0.3cm]
\reflionx & & \ion\tmark[c] & & $1.3^{+3.7}_{-0.3}$ \\
\\[-0.2cm]
& & $A_{\rm Fe}$\tmark[d] & & $>8.2$ \\
\\[-0.3cm]
\hline
\hline
\\[-0.3cm]
\rchi$(\mathrm{d.o.f})$ & & & & 1.1(466) \\
\\[-0.3cm]
$L_{0.3-10}$\tmark[e] & & & & $24.6^{+1.6}_{-6.7}$ \\
\\[-0.3cm]
\hline
\hline
\end{tabular}
\label{tab_refl}
\end{center}
\small $^a$ Column densities are given in $10^{21}~$\atpcm \\
\small $^b$ Radii are given in units of gravitational radii, \rg\ = $GM_{BH}/c^{2}$ \\
\small $^c$ Ionisation parameter, given in \ergcmps \\
\small $^d$ iron abundance, quoted relative to the solar value \\
\small $^e$ The absorption corrected 0.3--10.0\,\kev\ luminosity in $10^{40}$ \ergps \\
\end{table}

\subsection{Comptonisation Interpretation}
\label{subsec_spec_compt}

\begin{figure*}
\begin{center}
\rotatebox{0}{
{\includegraphics[width=240pt]{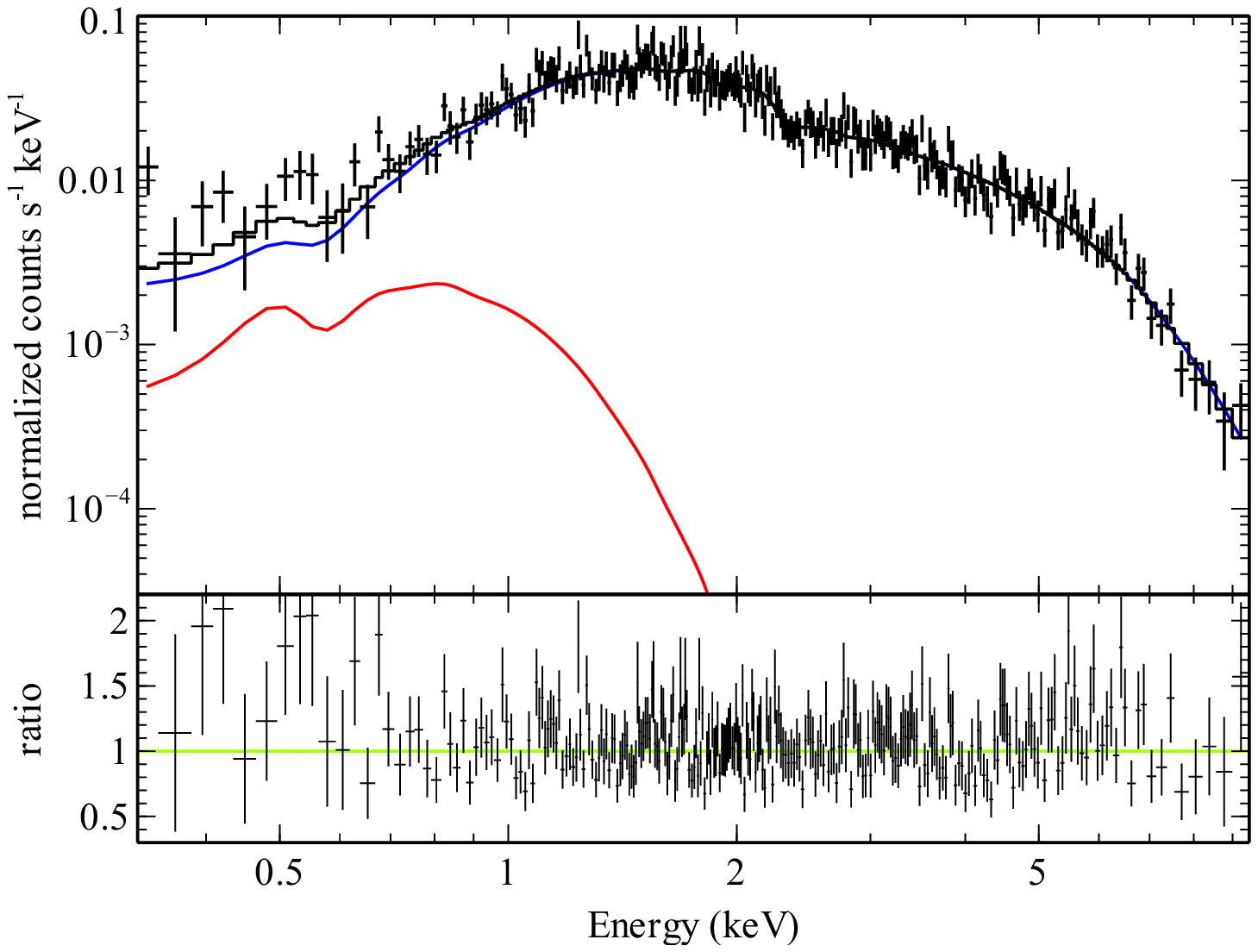}}
}
\rotatebox{0}{
{\includegraphics[width=240pt]{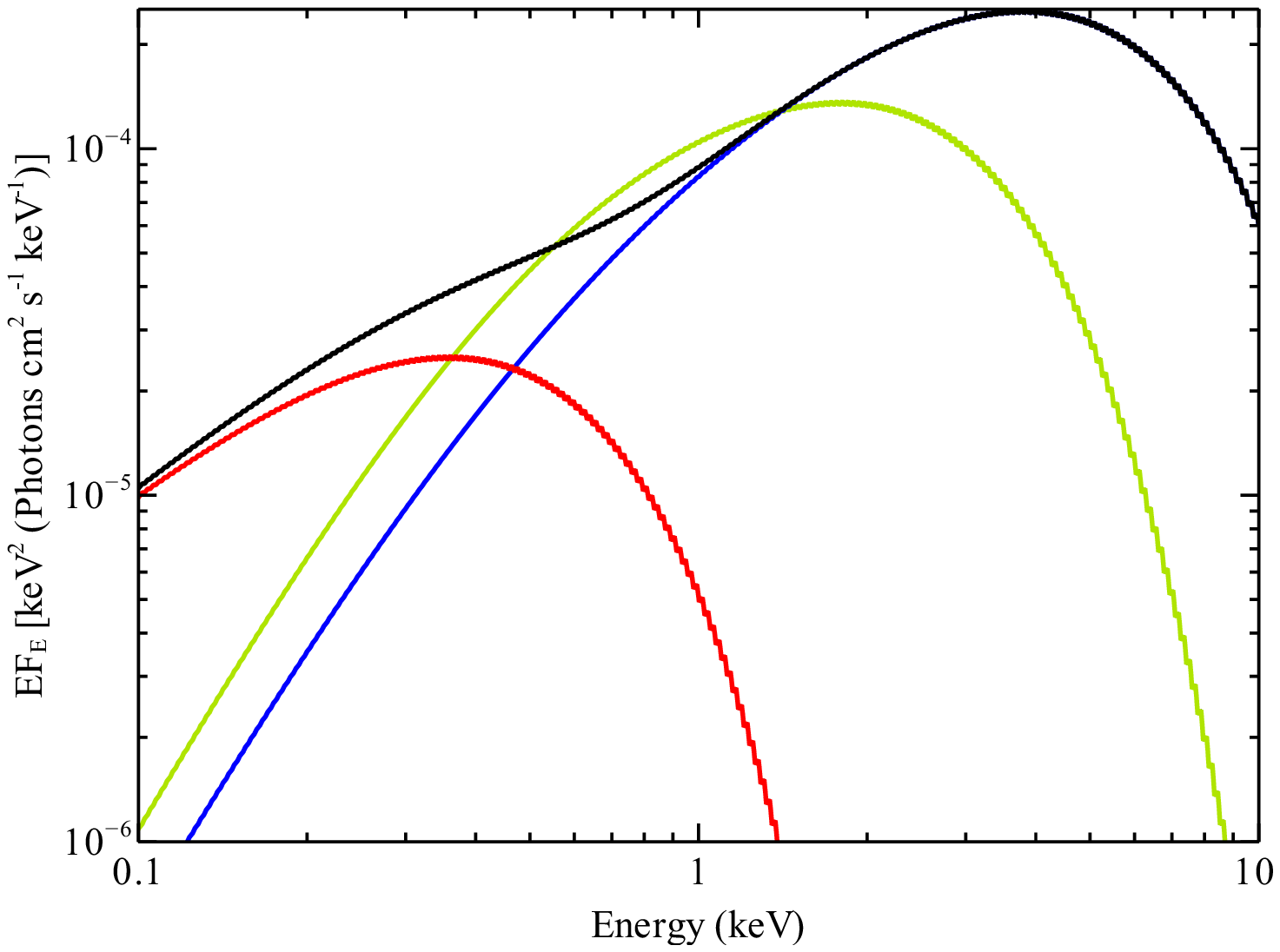}}
}
\end{center}
\caption{The Comptonisation model applied to the spectrum of \ulxone, and the relative
contributions of the components; the outer (observable) disc is shown in red and the
Comptonised emission in blue, while the overall model is in black. Again only the
\epicpn\ data are shown for clarity. The left panel shows the model fit to the photon
spectrum, while the right panel shows the proposed model (absorption corrected) and its
components in an $EF_E$ plot, highlighting the effects of the corona on the observed
disc emission; the additional emission from the disc that would be observed if the
corona and its effects were removed is shown in green.}
\label{fig_compt}
\end{figure*}

The second model applied is that of \cite{Gladstone09}, which proposes a new accretion
state for X-ray binaries, dubbed the `Ultraluminous State'. In this scenario, the
turnover is due to Comptonisation of disc photons by an optically thick corona. To model
this, we make use of the \dkbbfth\ code (\citealt{dkbbfth}), which accounts for the
energetic coupling between the disc and the corona using the model of \cite{Svensson94},
\ie it assumes the corona extends homogeneously (constant temperature and optical depth)
over the disc out to some transition radius, $R_{\rm c}$. The corona takes some fraction,
$f$, of the gravitational energy available within $R_{\rm c}$, with the remaining fraction
$(1-f)$ powering the inner disc emission. As a simplifying assumption the model also
takes $f$ to be constant with radius. Other authors have proposed alternative scenarios
in which $f$ may vary with radius (\eg \citealt{Janiuk00}), but as shown in
\cite{dkbbfth}, keeping $f$ constant is an adequate approximation for even the best
quality XRB data, and so will be sufficient for our needs. The key parameters of
\dkbbfth\ are the temperature of the Comptonising electrons, $T_{\rm e}$, the spectral
index of the corona, $\Gamma$, the transition radius (or the outer radius of the corona),
and the inner temperature of the accretion disc, $T_{\rm in}$. Note that this temperature
is what would be observed \textit{if the corona and its effects on the disc were
completely removed}. The fraction of the power in the corona $f$ is calculated self
consistently via iteration from the coronal spectral parameters. The inner radius of the
disc, \rin, may be calculated from the normalisation of the model if the distance to
the source and the inclination of the (inner) disc are known, in the same manner as for
the \diskbb model, and the optical depth, $\tau$, can be calculated from $\Gamma$ and
$T_{\rm e}$ (assuming a slab geometry).

We applied this model to the data, allowing for both Galactic and intrinsic neutral
absorption in the same way as for the reflection interpretation, using \phabs and
\zphabs again, and obtained a good fit with \rchi$(\mathrm{d.o.f})$ = 1.1(471). Model
parameters, and key quantities derived from them, are given in Table \ref{tab_compt},
and the fit to the data are shown in Fig. \ref{fig_compt}. As with the reflection
model, we find similar parameters here to those obtained in the original work of
\cite{Gladstone09}. The electron temperature of the Comptonising corona is low
($kT_{\rm e} \sim 1.4$~\kev), and it is optically thick ($\tau \sim 9$). In addition,
the temperature of the innermost disc inferred from the model ($kT_{\rm in} \sim 0.9$~\kev,
in the limit of no corona) is akin with that seen in stellar mass black hole binaries
in accretion states with high Eddington ratios, \eg the high/soft state or the steep
powerlaw state (using the nomenclature of \citealt{Remillard06}). The inclination angle of
the inner disc, $i$, is not known in this case, so the inner radius of the disc is
quoted in terms of its functional dependence on this angle.

\begin{table}
  \caption{Parameters obtained, and quantities derived from them, with the Comptonisation
interpretation for the spectrum of NGC 4517 ULX1; parameters marked with * have not been
allowed to vary.}
\begin{center}
\begin{tabular}{c p{0.2cm} c p{0.2cm} c}
\hline
\hline
\\[-0.3cm]
Component & & Parameter & & Value \\
\\[-0.3cm]
\hline
\hline
\\[-0.3cm]
\phabs & & \nh\tmark[a] & & $0.188$* \\
\\[-0.3cm]
\hline
\\[-0.3cm]
\zphabs & & \nh\tmark[a] & & $4.2^{+0.4}_{-0.3}$ \\
\\[-0.2cm]
& & $z$ & & $0.003764$* \\
\\[-0.3cm]
\hline
\\[-0.3cm]
\dkbbfth & & $\Gamma$ & & $1.76^{+0.09}_{-0.06}$ \\
\\[-0.2cm]
& & $kT_{\rm e}$\tmark[b] & & $1.4^{+0.2}_{-0.1}$ \\
\\[-0.2cm]
& & $kT_{\rm in}$\tmark[b] & & $0.9^{+0.3}_{-0.2}$ \\
\\[-0.2cm]
& & $R_{\rm in}\sqrt{{\rm cos}~i}$ \tmark[c] & & $390^{+700}_{-250}$ \\
\\[-0.2cm]
& & $R_{\rm c}$/\rin\tmark[d] & & $10.7^{+5.7}_{-3.3}$ \\
\\[-0.2cm]
& & $\tau$\tmark[e] & & $9.04^{+0.09}_{-0.06}$ \\
\\[-0.3cm]
\hline
\hline
\\[-0.3cm]
\rchi$(\mathrm{d.o.f})$ & & & & 1.1(471) \\
\\[-0.3cm]
$L_{0.3-10}$\tmark[f] & & & & $3.53 \pm 0.05$ \\
\\[-0.3cm]
\hline
\hline
\end{tabular}
\label{tab_compt}
\end{center}
\small $^a$ Column densities are given in $10^{21}~$\atpcm \\
\small $^b$ Temperatures are given in \kev \\
\small $^c$ Inner radius of the disc in km, in terms of the unknown inclination angle \\
\small $^d$ Outer radius of the corona, relative to the inner radius of the disc \\
\small $^e$ $\tau$ is not a model parameter, but is calculated from $\Gamma$ and $T_{e}$ \\
\small $^f$ The absorption corrected 0.3--10.0\,\kev\ luminosity in $10^{40}$ \ergps
\end{table}

\section{Discussion and Comparison}
\label{sec_dis}

We find that both disc reflection and Comptonisation can model the high quality
0.3--10.0\,\kev\ EPIC spectra of \ulxone\ well when a broad parameter range is
considered, and in fact give statistically equivalent fits to the data. This
appears to be the case for the majority of ULXs with spectral curvature, as all
the sources modelled with reflection in \cite{Caball10} are also well modelled
with Comptonisation by \cite{Gladstone09}. Currently these spectra represent the
best available ULX data, so other means are necessary to distinguish between
these two interpretations. We begin by considering the physical implications of
our analysis, as well as the results obtained for other ULXs.

The lack of direct thermal emission from the accretion disc included in the reflection
model may at first seem puzzling. For many of the black hole masses invoked in both
the stellar mass \textit{and} IMBH interpretations, the inner temperature of the disc
is expected to occur in the X-ray bandpass ($\sim$1.0\,\kev\ for a $\sim$10\,$\msun$
stellar mass black hole, down to $\sim$0.3\,\kev\ for a $\sim$1000\,$\msun$ IMBH).
Given the strong disc reflection, one might expect to see a sizeable contribution from
a direct thermal component. However, if magnetic extraction of energy from the disc to
the corona (which itself might be magnetic in nature, see \eg \citealt{Merloni02},
\citealt{Liu02} and references therein) is highly efficient, energy may be extracted
from the disc before it is able to thermalise. In this way, the direct thermal emission
could be suppressed with respect to that of the corona, and hence also the reflected
component. Such processes are not currently well understood and merit further attention.
Were this to be the case, such highly efficient magnetic extraction would distinguish
ULXs from both XRBs and AGNs, as it is well known that there \textit{is} often evidence
for direct thermal emission in both cases (albeit in the UV for AGNs), especially at
high accretion rates (as are anticipated to be present in ULXs). Hence the efficiency
of magnetic extraction is not expected to be particularly high in XRBs and AGNs.

Of course, the direct thermal disc emission must be distinguished from any of the reflected
emission thermalised in the disc during the reflection process. The temperature at which
this feature appears, for a given ionisation state, depends on the rate of free-free
absorption and hence on the density of the illuminated surface of the disc. It is
incorporated into the \reflionx\ model, however, this model is calculated assuming a
hydrogen number density of $n_{\rm H} = 10^{15}$ cm$^{-3}$, for which the thermalised
component of the reflected emission occurs at $\lesssim$0.1\,\kev; if the assumed density
is not appropriate for the discs around ULXs then this emission will not be accounted for
correctly. For \ulxone\ this might not be a significant issue as the large absorption
column prevents a detailed study of any additional emission at soft X-ray energies. For
sources with less absorption it will be necessary to include this effect with reflection
models calculated for a higher density (\eg \refhiden; \citealt{refhiden}). This will be
explored in future work.

The emissivity profile obtained is strongly centrally peaked, which suggests that
gravitational light bending is important. For this to be the case, the corona must be
compact and located close enough to the central black hole that the gravitational
potential focuses additional flux from the corona away from the observer and onto the
the accretion disc. This allows the reflected component to appear more luminous than
the intrinsic component; see \cite{lightbending} for a more detailed discussion of
this phenomenon. In their work, it is shown that gravitational light bending
preferentially illuminates the inner disc, and hence in such a regime the radial
emissivity profile of the inner disc is not best described by a single uniform
powerlaw, so the significant improvement obtained by allowing for a broken powerlaw
emissivity profile is not unexpected. Given that centrally peaked emissivity profiles
appear to occur fairly frequently when applying a disc reflection interpretation to
ULX spectra, and that some sources require reflection dominated solutions, allowing
for a more complex emissivity profile may lead to similar improvements in the
modelling of other sources.

The proximity of the X-ray emitting corona to the black hole required to provide the
obtained emissivity profile for \ulxone\ suggests that, for a corona on-axis, the
majority of the flux ends up crossing the event horizon rather than illuminating the
disc. The model of \cite{Nied08} requires the corona to be within $\sim$1.5\,\rg, at
which point, if on-axis, $\sim$90 per cent of the coronal flux may not illuminate the
disc, so the true intrinsic luminosity would be substantially larger than that
observed. In addition, any small change in the height of the corona above the black
hole would lead to large changes in observed luminosity, which may be in contrast with
the extremely stable ULX lightcurves seen here and in \cite{Heil09}. A more
appropriate geometry may be similar to that proposed for MCG --6-30-15 by \cite{Nied08},
where the corona is located just above the inner disc. In this scenario the illumination
from the corona is Doppler beamed along the disc, so a greater proportion illuminates
the disc rather than being lost over the event horizon.

At a first glance, the high iron abundance obtained for the accretion disc here,
and in general when modelling ULX spectra with disc reflection ({\citealt{Caball10}),
does not seem consistent with the observation that ULXs are often located in low
metallicity regions (see \eg \citealt{Soria05}, \citealt{Mapelli09}, \citealt{Mapelli10}),
nor the theoretical considerations that suggest the \mbh\,$>$10\,$\msun$ black holes ULXs
are speculated to host are easier to form in low metallicity regions, both due
to the formation of larger progenitor stars, and reduced mass loss rates from stellar
winds (\citealt{Madau01}; \citealt{Bromm04}; \citealt{Vink01}).
However, if the black holes in ULXs are accreting from debris discs formed from the metal
rich inner layers of the stars that originally collapsed to form them in the first place,
it may be possible for the accretion disc to have a very high metallicity even though
the surrounding regions do not. Such discs could be produced by a method similar
to that proposed by \cite{Li03}, where the inner regions of the collapsing/exploding star
are not properly ejected, and fall back towards the collapsed core to form the accretion
disc. In such a scenario the black hole need not be in a mass transfer binary system to
appear bright in X-rays, and we note that a number of ULX optical counterparts appear more
consistent with X-ray irradiation of the outer regions of an accretion disc than with a
stellar companion (\eg \citealt{Kaaret09}), although it may just be that the emission from
the disc dominates that of any companion star even in the optical bandpass, as is the case
for LMXBs in outburst. Debris discs might be a common scenario if ULXs are associated with
supernova remnants (SNR), an explanation initially proposed for the ionised nebulae many
ULXs are embedded within, \eg IC\thinspace342\thinspace X-1 (\citealt{Roberts03}) and
MF\thinspace16 in NGC\thinspace6946 (\citealt{RobCol03}).

However, there is growing evidence that these nebulae are not SNR, but a combination
of X-ray photoionised and shock-ionised material due to irradiation and outflowing
winds from the central X-ray source (see \eg \citealt{Pakull08}). From these nebulae
the ages of ULXs can be estimated as $\sim$10$^{4-6}$ yr (\citealt{Pakull06};
\citealt{Pakull08}), which can be used to estimate the mass of high metallicity
material required in the debris disc scenario. In order to radiate at \lx\,$>$\,$10^{39}$
\ergps\ the black hole must be accreting at \mdot\ $ > 10^{-7}$ \msun\ yr$^{-1}$, so
in order for ULXs to have observable debris discs, these must have formed from at
least $\sim$0.1 \msun\ of high metallicity material. Of course, this makes the
simplistic assumption that the mass accretion rate is constant with time.
\cite{Mineshige97} studied post supernova accretion via fallback of matter in detail
and found that the accretion rate should evolve with time, as \mdot\ $\propto t^{-a}$,
with $1 < a < 1.5$ (the exact value of $a$ depends on the conditions of the material
as it begins to fall back). In this case the required accretion rate of \mdot\ 
$ > 10^{-7}$ \msun\ yr$^{-1}$ would represent the `current' accretion rate, and the
initial mass of high metallicity material available may have had to have been much
greater than 0.1\,\msun\ even for a ULX currently radiating at $10^{39}$ \ergps. In
addition, disc winds may have been present throughout, or at stages during the lifetime
of the ULX. The initial disc masses required are therefore at least an order of
magnitude larger than the lower limits for the current disc mass estimates for the ULXs
CXOJ033831.8-352604 and XMMU122939.7+075333 in \cite{Porter10}.

It may be difficult to form such large discs through fallback. Supernovae in low
metallicity regions may not eject large amounts of material (\citealt{Zampieri09}), hence
massive, rapidly rotating progenitor stars may be required. In addition, a number of optical
ULX counterparts have been identified as probable high mass or evolved stellar companions,
(see \eg \citealt{Kuntz05}, \citealt{Soria05}, \citealt{Liu07}, \citealt{Roberts08}, but
especially \citealt{Patruno08}), from which mass \textit{is} expected to be transferred to
the black hole. The binary companion and the mass transfer could disrupt and dilute the
metallicity of such a large disc. The possible presence of outflows is also interesting.
The population of optical nebulae associated with ULXs appear to show a combination of
photoionised and shock-ionised material, and \cite{Pakull08} argue the latter cases may be
inflated by winds from the central source, although the possibility that they are remnants
of the explosive event in which the ULX was formed is not ruled out. If these are due to
winds and ULXs are accreting from high metallicity fallback material, some metal enrichment
of the regions surrounding the ULX might be expected, although exactly how much effect on
the elemental abundances these outflows should have will depend on the amount of material
ejected in comparison to that in the ISM. It seems that although debris discs can not be
ruled out for a fraction of the population, they may not be a viable explanation for all ULXs.

An alternative explanation is that the iron abundances obtained may not be good
estimates of the true abundances of the systems in question, but systematically
modified by processes that are not accounted for in the modelling. For example,
this would be the case if there are actually multiple reflections of the intrinsic
continuum. The observed emissivity profile
implies that strong gravitational light bending is an important process in \ulxone,
and ULXs in general, if reflection is the correct interpretation. This causes the
initial reflection of the intrinsic powerlaw component to occur primarily within
the central regions of the accretion disc, so the majority of the reflected
emission also originates within the region in which strong light bending is
present. Some fraction of this reflected emission could therefore be bent back
towards the disc and reflected a second time. The initially reflected emission
already has the iron emission features imprinted onto it and the secondary
reflection should serve to enhance these further, as demonstrated by \cite{Ross02},
hence the iron abundance obtained from modelling the emission with a single
reflection process could be artificially inflated. If this is the case, it may not
be necessary to invoke the high metallicity debris discs discussed previously.
Whatever the correct explanation for the super-solar iron abundances
obtained, it might be natural to expect that other heavy elements might be or
appear over-abundant for the same reason. Unfortunately model limitations currently
prevent us from exploring this possibility.

The Comptonisation interpretation requires that the coronal electron temperature of
\ulxone\ is rather low, at $kT_{\rm e} \sim 1$~\kev. Indeed, low electron temperatures
($kT_{\rm e} \sim 2$~\kev) are a common feature in the application of this model to ULX
spectra (\citealt{Gladstone09}), as they are an unavoidable consequence of modelling
the observed curvature as the high energy cut-off of Compton up-scattering by thermal
electrons. Typical electron temperatures for XRBs are usually much higher, $kT_{\rm e}
\sim 50-100$~\kev\ for binaries in the low/hard state and $\sim$30--50~\kev\ in the
high/soft state (although here there can also be a significant contribution from a
non-thermal electron distribution); see \eg \cite{Malzac09}, \cite{Gierlinski99},
\cite{Sunyaev80} amongst others. The coronal optical depth is also quite discrepant
from typical values for XRBs: $\tau \sim0.1-0.3$ in the high/soft state and $\sim$1
in the low/hard state. Here we find $\tau \sim 9$. However, ULX spectra are often
likened most to the steep powerlaw state. Sources in this state often display very
high Eddington Ratios ($L/L_{\rm E} \sim$1). Again, Compton up-scattering by thermal
electrons is observed to be important, with $kT_{\rm e} \sim10-30$~\kev\
(\citealt{Wilson01}; \citealt{Kubota04}; \citealt{KubDone04}). These temperatures
are often still highly discrepant from those obtained from ULX spectra, although
the highest luminosity observations of the XRB GRS\thinspace 1915+105 are notable
exceptions (\citealt{Middleton06}; \citealt{Ueda09}).

However, in the work of \cite{Gladstone09}, it is suggested that most ULXs may represent
a new accretion state entirely, dubbed the `Ultraluminous State', which sources enter
if and when they are able to radiate at super-Eddington luminosities ($L/L_{\rm E} > 1$). As 
sources approach, and potentially exceed the Eddington limit, there is much speculation
that radiatively driven disc winds become increasingly important (see \eg \citealt{King03}).
A case in point, potentially relevant for ULXs, is the binary system SS\thinspace433 which
is expected to have a very high mass transfer rate and from which large outflows are
observed, in addition to relativistic jets (\citealt{Begelman06}; \citealt{Poutanen07}).
\cite{Gladstone09} argue that such winds may be the source of the increased coronal
opacity obtained from ULX spectra, as further mass is injected into the corona by
the outflow. The electron density of the corona is then increased, and there is therefore
less energy available per electron, so the average temperature of the electrons
decreases. These outflows could be responsible for the inflation of the observed ULX optical
nebulae (\citealt{Pakull08}).

\begin{figure}
\begin{center}
\rotatebox{0}{
{\includegraphics[width=220pt]{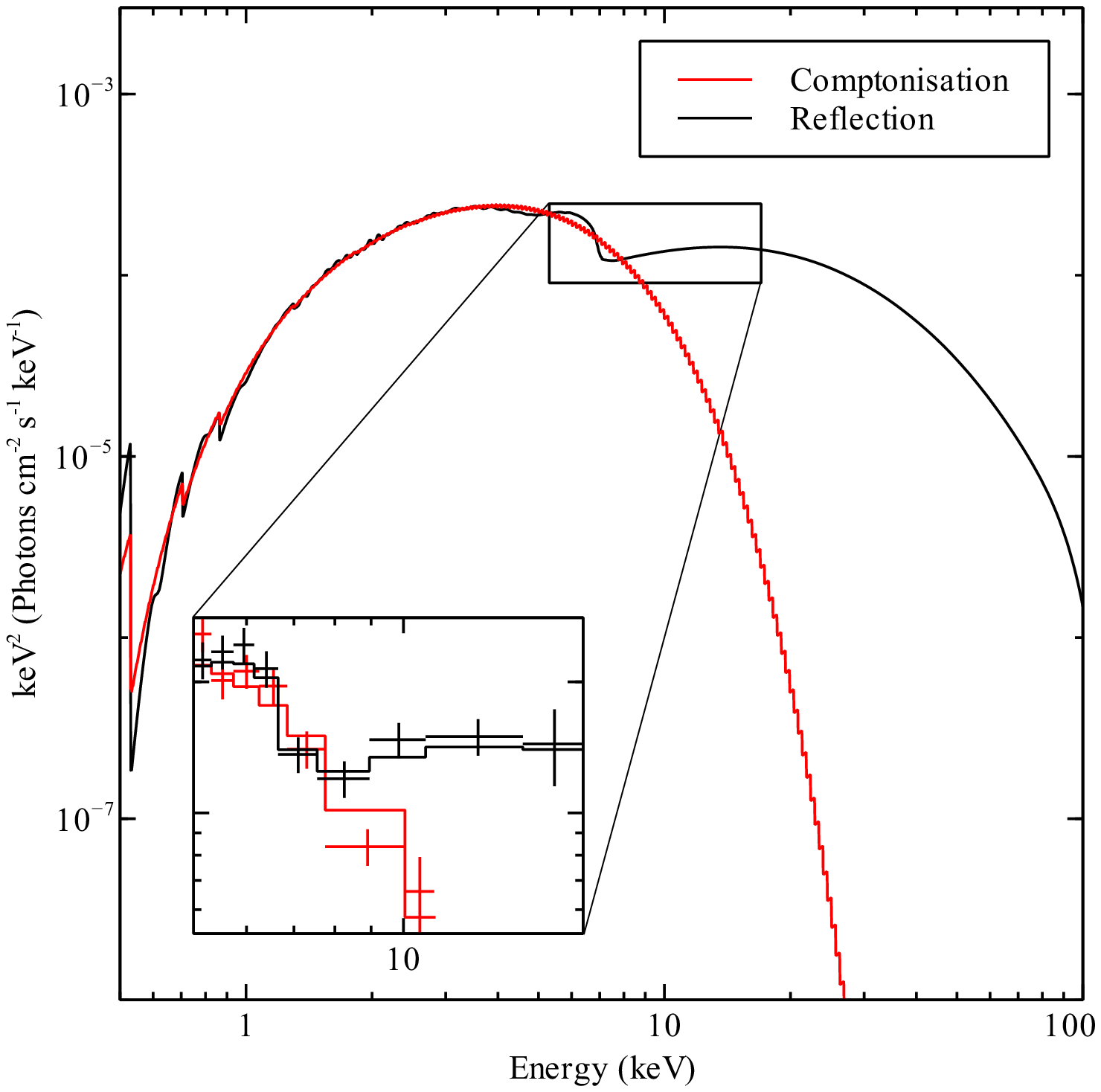}}
}
\end{center}
\caption{A comparison of the Comptonisation and disc reflection models for ULX
spectra applied to \ulxone; both have been extended beyond the \xmm bandpass up
to 100\kev. Although the models agree well below 10\kev, there is a clear
discrepancy of many orders of magnitude between them at higher energies, due to
the presence of the Compton hump in reflection spectra. Inset: data simulated
using the current \textit{Astro-H} response files for a 100\,ks observation
with both models. \ulxone\ is clearly detected to higher energies with the
reflection interpretation.}
\label{fig_comparison}
\end{figure}

In brief, it seems that on purely physical grounds, neither interpretation can be
excluded from being the underlying process for at least some fraction of the ULX
population, so we must seek other, observational means to distinguish between them.
Ideally, this would be easily applied to a sample of ULXs large enough to be
considered representative of the population as a whole. Unfortunately, the models
do not make unique predictions for the emission at non X-ray wavelengths, so
multiwavelength studies are unlikely to resolve this particular issue. In Fig.
\ref{fig_comparison} we compare the two models fit to the spectrum of \ulxone, but
extended beyond the \xmm bandpass up to 100\,\kev. A few minor details aside, the
models agree very well below $\sim$10\,\kev. As we have demonstrated an observation
of $\sim$100\,ks is insufficient to differentiate between them for \ulxone, and the
work of \cite{Caball10} and \cite{Gladstone09} suggests both models agree well at
these energies for similar length observations of even brighter ULXs. Attempting to
determine which, if any, of these models is correct using only CCD data would require
even longer observations, which are not realistically obtainable for a large number
of sources. Some ULXs may be bright enough that the reflection grating spectrometers
aboard \xmm could investigate the differences in the neutral absorption the two
models predict with reasonable observation lengths. However, these are very much in
the minority, and obtaining an independent value for the column density would not
confirm either interpretation directly, it would merely provide an an additional
constraint that each model must fulfil if they are to continue to provide acceptable
representations of the data.

Potentially, the most fruitful avenue utilising only data below 10\,\kev\ is the
combination of spectral and variability studies. Investigation of the rms variability
spectrum can aid significantly in the deconvolution of separately varying spectral
components and could be able to differentiate between the two models considered here.
Both are two component solutions, but the energy dependence of the relative importance
of the two components is different for the two models. This approach is being
investigated by Middleton et al. (in preparation). However, a reliable estimation of
the rms variability spectrum requires good quality data, which often requires long
observations. In addition, this method relies on the ULXs displaying observable
variability, however \cite{Heil09} demonstrated that the brightness of many ULXs is
remarkably stable within single \xmm observations up to $\sim$100\,ks in duration;
it seems that \ulxone\ may also be counted amongst this number. As a further
complication, there are also sources that exhibit flux variability without any strong
spectral variability (see \citealt{Kaaret09}.) This might be considered unusual for
interpretations invoking two distinct emission components. However, these components are
not independent of one another, in both models one component represents reprocessed
emission from the other. If the luminosity of the primary component varies, then the
luminosity of the secondary component must also vary in the same way. A brightening
of the primary component might lead to some variation in the observable parameters,
for example the temperatures of the observable part of the disc and the Comptonising
electrons in the Comptonisation interpretation or the ionisation of the disc in the
reflection interpretation, but as argued by \cite{Vierdayanti10} these effects are
likely to be fairly subtle and not easily detected with the hardness ratio analysis
employed by \cite{Kaaret09}. Unfortunately, combining spectral and variability studies
to decompose emission components is often only possible when strong spectral variability
is present. All of these requirements limit the number of sources for which such
analysis is possible.

Returning to Fig. \ref{fig_comparison}, the comparison above $\sim$10\,\kev\ shows the
two models diverging. The Comptonisation model continues its downward curvature, as this
is due to the high energy limit of the Compton up-scattering process of disc photons by
cool electrons, while the reflection spectrum turns back up above 10\,\kev\ due to the
Compton hump. This is a broad emission feature present in reflection spectra that peaks at
$\sim$20--50\,\kev\ due to the interplay within the reflecting medium between Compton
down-scattering of high energy photons and photoelectric absorption of low energy photons
(\citealt{George91}). Features consistent with the Compton hump are fairly commonly
observed in the spectra of both BHBs and AGNs (see \eg \citealt{Reis10lhs, Walton10Hex}).
The difference in the origins of the observed curvature between the two models leads to
large differences in their predicted high energy spectra. Of course, the Comptonisation
model assumes the electrons have a purely thermal distribution. It is likely there is an
additional non-thermal electron population that will eventually arrest the downward
curvature predicted in this Comptonisation model, but we still anticipate clear differences
between the two models at high energies, as for any reasonable non-thermal temperature
distribution these electrons would merely limit the downward curvature and provide a high
energy powerlaw tail, not turn the spectrum back up again. In addition, if ULXs are an
extremely high accretion rate extension of the Steep Powerlaw State (\ie the Ultraluminous
State), the emission from the non-thermal electrons may not be significant until energies
$\gtrsim100$\,\kev.

The strength of the predicted Compton hump is obviously determined by the parameters of the
reflection model, notably the inclination of the disc. Although we find the key parameters
to be well determined in this case, parameter degeneracies may exist in other cases.
\reflionx\ is an angle averaged model, best suited for use with inclination angles
$\sim$45\deg, so we investigate how the strength of the Compton hump may change for extreme
inclinations with the \pexrav\ model (\citealt{pexrav}). At low inclinations (10\deg)
its strength is actually enhanced by $\sim$5--10 per cent (relative to the 45\deg\ case),
while at high inclinations (80\deg) its strength is reduced by $\sim$25--30 per cent. This
variation is highly unlikely to be able to reconcile the predictions of the two models in
any given case. Other factors may also modify its strength. For example, \reflionx\ is
calculated assuming the illuminating spectrum is a powerlaw with a high energy cut-off at
300\,\kev. Lowering the energy of this cut-off will also serve to reduce the predicted Compton
hump to some extent. However, we stress that when both models provide acceptable representations
of the data below 10\,\kev, the extra emission due to the presence of this high energy feature in
the reflection interpretation should lead to large differences in the predictions of the two
models above 10\,\kev\ for just about any parameter combination. Therefore, observations of ULXs
at such energies should be an extremely important diagnostic of the underlying physical processes
powering the emission from ULXs.

At the time of writing there are two facilities that are capable of observing
\textit{simultaneously} in the $\sim$0.5--10.0\,\kev\ range required to determine
the key model parameters for each interpretation from the curvature at $\sim$6\,\kev, and
at the higher energies required to distinguish between them. These are \suzaku\
(\citealt{SUZAKU}), with the combination of the Hard X-ray Detector (HXD, specifically the
PIN detector; \citealt{SUZAKU_HXD}) and X-ray Imaging Spectrometers (XIS; \citealt{SUZAKU_XIS}),
and \swift\ (\citealt{SWIFT}) with the combination of the Burst Alert Telescope (BAT;
\citealt{SWIFT_BAT}) and the X-ray Telescope (XRT; \citealt{SWIFT_XRT}). Some key
information on the hard X-ray instruments on board these two satellites is summarised in
Table \ref{tab_det}. Both have very modest spatial resolution (the HXD is a collimating
instrument, so its effective spatial resolution is the same as its field of view). For
the vast majority of ULXs, and certainly for the specific case of \ulxone\ we have been
considering here, neither of these instruments would be able to resolve the ULX from the
nucleus of its host galaxy, which in most galaxies (although not necessarily \ngc4517)
will be the dominant source of hard X-rays. This has been a common problem with hard
X-ray detectors to date, and as such very little is known about the spectra of
ULXs above 10\,\kev.

\begin{table}
  \caption{Key information for the Hard X-ray Detector (HXD) and the Burst Alert
Telescope (BAT) aboard the \suzaku\ and \swift\ satellites respectively.}
\begin{center}
\begin{tabular}{c c c c}
\hline
\hline
\\[-0.3cm]
Instrument & Type & Energy Range & Angular Resolution \\
\\[-0.3cm]
& & (\kev) & (FWHM, arcmin) \\
\\[-0.3cm]
\hline
\\[-0.3cm]
HXD (PIN) & Collimating & 12--70 & 34 (FoV) \\
\\[-0.3cm]
BAT & Coded Mask & 15--150 & 17 (PSF) \\
\\[-0.3cm]
\hline
\hline
\end{tabular}
\label{tab_det}
\end{center}
\end{table}

The exception is M\,82~X-1, for which \cite{Miyawaki09} claim a \suzaku\ HXD
detection up to $\sim$20\,\kev. In this work, the 3.2--20.0\,\kev\ spectrum of M\,82~X-1
is found to require a curved continuum (below $\sim$3.2\,\kev\ the spectrum is dominated
by diffuse, thermal emission). Some uncertainty remains over what proportion of the flux
detected above 10\,\kev\ originates from M\,82~X-1 and what proportion originates from
the rest of the (sizeable) XRB/ULX population of M\,82, but the requirement of a curved
continuum is not sensitive to this. This curvature is well modelled with Comptonisation,
and similar parameters to the results from other ULXs are obtained ($kT_{\rm e} \sim$
2.5\,\kev, $\tau \sim$ 8). However, this was only possible due to the unique scenario
M\,82~X-1 is in. M\,82 is a starburst galaxy with no discernible AGN activity, and of its
X-ray source population, M\,82~X-1 is usually the brightest (\citealt{Zhang09};
\citealt{Zezas01}; \citealt{Matsumoto01}). In addition, during the observations utilised
by \cite{Miyawaki09}, it was possible to choose the pointing to ensure other nearby bright
X-ray sources, \eg the LLAGN in M\,81, M\,81~X-6 and Holmberg\,IX~X-1, were outside of
the HXD field of view.

It may be possible for the PIN and BAT detectors to contribute to the differentiation of
other sources if they are also significantly brighter than any others that they can not
spatially be resolved from. In addition, a total non-detection at a low enough flux
sensitivity would rule out the reflection interpretation. However, the sensitivities of
these instruments is rather limiting for the study of ULXs. We stress that these
scenarios will both be extremely rare, so such an approach will not be possible for the
vast majority of the ULX population. M\,82~X-1 is the only currently known source these
instruments have been able to study due to its apparently unique situation and high flux.
In order to obtain spectral information above 10\,\kev\ for a large sample of ULXs an
imaging spectrometer with significantly higher spatial resolution than is currently
available will be required. With the improvement in sensitivity and angular resolution
($\sim$60--90'') offered by the Hard X-ray Imaging System proposed to fly on
\textit{Astro-H}\footnote{http://astro-h.isas.jaxa.jp/si/index\_e.html}
(\citealt{ASTROH_tmp}) in 2014, combined with the Soft X-ray Imaging system and the Soft
X-ray Spectroscopy System (microcalorimeter) also due to be on board, it should be
possible to begin addressing the nature of the observed emission from ULXs with a series
of relatively short, pointed observations. In the shorter term, such observations should
also be possible with simultaneous observations from \nustar\footnote{http://by134w.bay134.mail.live.com/default.aspx?wa=wsignin1.0}
(\citealt{NUSTAR_tmp}), which is due for launch in 2011, and either \xmm or \suzaku.
An excellent candidate with which to demonstrate the capabilities of these instruments
to contribute to ULX studies is Holmberg~IX X-1, which is bright, nearby, and has a
very hard spectrum, but currently cannot be resolved from the nucleus of M\thinspace81
above 10\,\kev.

\begin{figure}
\begin{center}
\rotatebox{0}{
{\includegraphics[width=238pt]{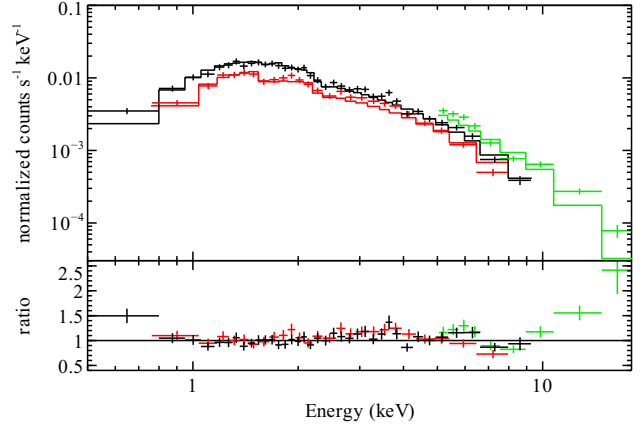}}
}
\end{center}
\caption{Data simulated from the reflection interpretation presented for \ulxone\ for the
suite of instrumentation aboard the \textit{Astro-H} satellite (black: soft X-ray
imaging system; red: X-ray microcalorimeter; green: hard X-ray imaging system). 
The solid lines show the application of the Comptonisation model to this simulated
data. Clearly there is an excess of counts over the model prediction at energies
above $\sim$10\,\kev.}
\label{fig_model_test}
\end{figure}

As an immediate demonstration of the potential of these two instruments for ULX
science, we have obtained the pre-flight instrument response files for \textit{Astro-H}
and simulated the data that would be obtained for both the models obtained here with a
100\,ks observation of \ulxone\ (we only simulate for \textit{Astro-H} because
the hard X-ray imaging responses are very similar for this mission and \nustar, and
\textit{Astro-H} will also carry soft X-ray spectrographs). These simulations are shown
in the inset in Fig. \ref{fig_comparison}. \ulxone\ would clearly be detected
out to much higher energies with the reflection interpretation than with the
Comptonisation interpretation. We apply the Comptonisation interpretation to the
data simulated from the reflection model for the full suite of instrumentation aboard
\textit{Astro-H}, allowing its parameters to vary (although requiring that Comptonisation
models the $\sim$6\,\kev\ curvature), and obtain a relatively poor fit,
with \rchi$(\mathrm{d.o.f})$ = 1.33(284). Even though the electron temperature of the
corona increases to $T_{\rm e} \sim 3.6$\,\kev\ and its optical depth decreases to $\tau \sim
3.5$, Fig. \ref{fig_model_test} shows there is a clear excess of counts at hard energies.
To provide a rough estimate of the number of ULXs for which \textit{Astro-H} and
\nustar\ should be able to make such a distinction with observations of a similar or
shorter duration to that considered above, there are 19 sources in the ULX catalogue
presented by Walton \etal (2011; submitted) with a greater flux than \ulxone,
including 5 whose flux is more than an order of magnitude higher. We stress that
this is actually a lower limit, as there are a number of bright ULXs that are not
present in the catalogue, of which Holmberg~IX X-1 is a prime example.

\section{Conclusions}
\label{sec_conc}

We have analysed the high quality EPIC spectrum of \ulxone, demonstrating that this source
exhibits spectral curvature at $\sim$6\,\kev, similar to that seen in the highest quality
observations of other ULXs to date. This feature has been modelled with both the recent
reflection and Comptonisation interpretations (see \citealt{Caball10} and
\citealt{Gladstone09} respectively) in an attempt to provide a direct comparison of the two.
The reflection interpretation requires that the majority of the emission occurs within the
innermost regions of the accretion disc. We find that the radial emissivity profile must be
strongly centrally peaked in order to provide the smooth downturn, which can be achieved if
the disc-corona geometry is such that strong gravitational light bending occurs
(\citealt{lightbending}). In order to model the curvature in \ulxone, the disc must have, or
appear to have, a highly super-solar iron abundance. This may be possible if the black hole
is accreting from a debris disc formed primarily from the metal-rich inner layers of the
star that originally collapsed to form it, or if the light bending is so strong that multiple
reflections of the intrinsic continuum occur. Meanwhile, we find that the Comptonisation
interpretation requires that the electron temperature of the Comptonising corona is very low,
and that the corona is optically thick. It is the low temperature of the thermal component of
the Comptonising electrons that provides the curvature in the spectrum, as this determines
the high energy cut-off of the Compton up-scattering process. These low temperatures and high
optical depths may be due to mass loading of the corona if disc winds are important at high
accretion rates, as proposed by \cite{King03}.

Neither model may be excluded physically, so alternative observational means are required
to distinguish between them. Correctly identifying the origins of the high energy curvature
could be a very important step in enhancing our understanding of ULXs, as it is one of their
characteristics that distinguishes them from XRBs. By extending both of the models to
energies above 10\,\kev, it becomes clear that there are large differences between them at
these energies (Fig. \ref{fig_comparison}). We expect this to be the case even if, as is
likely, there is a contribution from a non-thermal distribution of electrons to limit the
downward curvature in the Comptonisation model. The major cause of the discrepancy between
the two models at high energies is the presence of the Compton hump in reflection spectra,
a broad emission feature that often peaks at $\sim$30--50\,\kev. Observations of ULXs above
10\,\kev\ should provide an important diagnostic for the relative importance of these
physical processes in the ULX population.

Unfortunately, the angular resolution of the hard X-ray instruments aboard currently
operational facilities that \textit{simultaneously} provide spectral coverage over the
$\sim$0.5--10.0 and $\sim$10--100\,\kev\ energy ranges is not sufficient to provide
uncontaminated data for a large number of ULXs. We stress the importance of modelling
data above and below 10\,\kev\ obtained simultaneously as spectral variability and
accretion state transitions could otherwise lead to incorrect conclusions, especially
if ULXs do primarily represent a new accretion state. Hard X-ray imaging spectrometers
with good angular resolution, such as the \textit{Astro-H} Hard X-ray Imaging System,
should allow a good number of reliable ULX spectra above 10\,\kev\ to be obtained in
the future.

\section*{ACKNOWLEDGEMENTS}

DJW acknowledges the financial support provided by STFC, and ACF thanks the
Royal Society. Some of the figures included in this work have been produced
with the Veusz\footnote{http://home.gna.org/veusz/} plotting package, written
by Jeremy Sanders. The authors would also like to thank the anonymous referee
for useful comments and suggestions, and their attention to detail. This work
is based on \xmm observations, an ESA mission with instruments and
contributions directly funded by ESA member states and the USA (NASA). In
addition, this research has made use of the NASA/IPAC Extragalactic Database
(NED), operated by the Jet Propulsion Laboratory, California Institute of
Technology, as well as the Digitised Sky Survey (DSS), produced at the Space
Telescope Science Institute under U.S. Government grant NAG W-2166.

\bibliographystyle{mnras}
\bibliography{/home/dwalton/papers/references}

\label{lastpage}

\end{document}